\documentclass[fleqn,twoside]{article}
\usepackage{espcrc2}

\usepackage{graphicx}

\def\Sr2RuO4{Sr$_2$RuO$_4$}

\newcommand{\AmS}{{\protect\the\textfont2
  A\kern-.1667em\lower.5ex\hbox{M}\kern-.125emS}}

\title{Polarization analysis of the inelastic magnetic scattering in \Sr2RuO4}
\author{B. F\aa k
\address[CEA]{DSM/DRFMC/SPSMS, CEA Grenoble, 38054 Grenoble, France}
\address[ISIS]{ISIS Facility, Rutherford Appleton
Laboratory, Oxon OX11 0QX, England},
S. Raymond\addressmark[CEA],
F. Servant\address[CNRS]{Centre de Recherches sur les Tr\`es Basses Temp\'eratures, 
CNRS, BP 166, 38042 Grenoble, France},
P. Lejay\addressmark[CNRS] and 
J. Flouquet\addressmark[CEA]}

\begin{document}

\begin{abstract}
The spin fluctuations in the normal state of the unconventional superconductor \Sr2RuO4\ have been studied using inelastic neutron scattering with polarization analysis on single crystals. We find that the spin fluctuations are anisotropic with $\chi_c/\chi_{ab}$=2.0$\pm$0.4. No evidence for $Q$-independent or nearly-ferromagnetic spin fluctuations are found in the energy range 4--22 meV, in contradiction with NMR measurements.
\vspace{1pc}
\end{abstract}

\maketitle

\Sr2RuO4\ is a layered perovskite oxide that is isostructural with the high-$T_c$ material 
La$_{2-x}$Sr$_x$CuO$_4$. Below $T_c$=1.5 K,  \Sr2RuO4\ enters an unconventional 
superconducting state with spin-triplet pairing, possibly mediated by spin fluctuations.
The low-temperature normal state is a quasi two-dimensional Fermi liquid 
with enhanced-mass quasiparticles (see Ref. \cite{mackeno} for a recent review). 
Neutron scattering experiments show strong 
inelastic magnetic intensity due to two-dimensional spin fluctuations at the 
incommensurate (IC) position {\bf k}$_0$={\bf Q}$_{hk}$=(0.3,0.3), which is related to the nesting 
properties of the Fermi surface \cite{sidis,servant00,servant02,braden}. 
The relatively high intensity observed between 
the IC peaks could be due to $Q$-independent ("local") spin fluctuations, 
as suggested by NMR measurements \cite{ishida}. In order to determine whether this scattering 
is of magnetic origin, and also to study the anisotropy of the spin fluctuations at {\bf k}$_0$, 
we have made polarized inelastic neutron scattering measurements on single 
crystalline \Sr2RuO4.
 
Three high-quality single crystals of \Sr2RuO4\ of a total volume of 0.6 cm$^3$, already used in previous 
measurements \cite{servant00,servant02}, were aligned with the $a$--$b$ plane 
as scattering plane in an orange cryostat held at $T$=1.5 K, ie. all measurements were 
performed in the normal state. The inelastic neutron scattering measurements
were performed on  the thermal triple-axis spectrometer IN22 at the ILL using polarizing 
Heusler crystals as monochromator and analyzer. An energy resolution of 0.9 meV was 
obtained with a final energy of 14.7 meV. A PG filter in the scattered beam suppressed higher-order contamination. A spin flipper in the scattered beam and a Helmholtz coil around the sample allowed measurements of the spin-flip (SF) and non spin-flip (NSF) cross sections for neutrons polarized along the wave vector transfer {\bf Q}, $\sigma_x$, perpendicular to {\bf Q} in the $a$--$b$ plane, $\sigma_y$, and perpendicular to the scattering plane, $\sigma_z$. The flipping ratio for nuclear Bragg peaks was 19, implying that no flipping-ratio corrections are necessary. Combination of the six different cross sections allowed to determine the components of the magnetic contribution to the total scattering that corresponds to fluctuations in the $a$--$b$ plane, 
\begin{equation}
\chi_{ab}=\sigma^{SF}_x-\sigma^{SF}_y=\sigma^{NSF}_y-\sigma^{NSF}_x,
\end{equation}
and along the $c$ axis,
\begin{equation}
\chi_c=\sigma^{SF}_x-\sigma^{SF}_z=\sigma^{NSF}_z-\sigma^{NSF}_x.
\end{equation}
We have here assumed that the magnetic susceptibility is isotropic in the $a$--$b$ plane, i.e. $\chi_{ab}$=$\chi_a$=$\chi_b$.  
Due to the reduced intensity in polarized neutron experiments, a limited number of {\bf Q}'s and energies were studied.  

\begin{figure}[t]
\centering
\includegraphics[width=70mm]{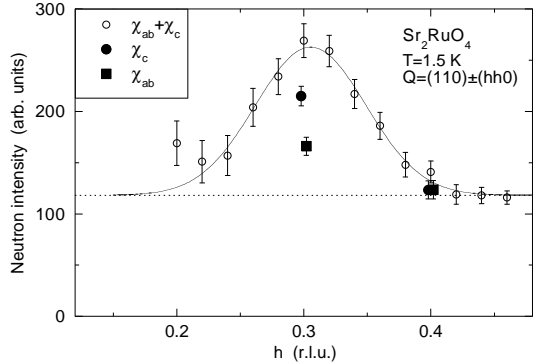}
\caption{\small Wave-vector dependence of the incommensurate spin fluctuations at $E$=6 meV. Open circles are measured near  {\bf Q}=(0.7,0.7,0) \cite{servant00,servant02} using unpolarized neutrons, which probes $\chi_{ab}$+$\chi_c$+non-magnetic scattering. 
Closed symbols are measured  near {\bf Q}=(0.3,0.3,0) using polarized neutrons, where  $\chi_{ab}$ and $\chi_c$ are probed separately. In order to obtain the same intensity scale, the polarized  data are rescaled with the magnetic form factor and a constant of 118, corresponding to the non-magnetic scattering, is added. }
\label{Qdep}
\end{figure}
\begin{table}[b]
\caption{\small Magnetic intensity polarized along the $c$ and $a$ axis (arb. units) at $E$=6 meV for two different wave vectors {\bf Q}.}
\begin{tabular}{ccc}
\hline
          		& {\bf Q}=(0.3,0.3,0) & {\bf Q}=(0.4,0.4,0) \\
\hline
$\chi_c$                 & 48.8 $\pm$ 4.8 & 2.7 $\pm$ 4.4 \\
$\chi_{ab}$                 & 24.0 $\pm$ 4.4 & 2.8 $\pm$ 4.5 \\
\hline
\end{tabular}\\[2pt]
\label{TabMag}
\end{table}

Figure \ref{Qdep} shows the spin-fluctuation response centered at the incommensurate position  {\bf k}$_0$ at an energy transfer of 6 meV using unpolarized neutrons \cite{servant00}, which probes $\chi_c$+$\chi_{ab}$. It has been speculated that the "background", i.e. the relatively high intensity level under the peak, could be due to $Q$-independent (local) spin fluctuations. Polarization analysis  provides unambiguous information on whether a signal is of magnetic or nuclear origin. The results of the present measurements are shown in Table \ref{TabMag} 
and Fig. \ref{Qdep}. 

Two conclusions can be drawn from Table \ref{TabMag}. ~(i) The "background" measured at {\bf Q}=(0.4,0.4,0)  is of non-magnetic origin, i.e. there is no evidence for local spin fluctuations at this energy. 
~(ii) The spin fluctuations at {\bf Q}=(0.3,0.3,0) are anisotropic, with $\chi_c/\chi_{ab}$=2.0$\pm$0.4. Measurements at a larger wave vector,  {\bf Q}=(0.3,0.7,0), confirm these results. Polarized work at even larger wave vectors, such as {\bf Q}=(0.7,0.7,0), are difficult due to the strong $Q$ dependence of the magnetic form factor. 

In earlier neutron scattering work, the indication of isotropic spin fluctuations were based on a study of the {\bf Q} dependence of the magnetic signal  \cite{servant02}. 
However, this method is easily affected by systematic effects such as changes in the orientation of the resolution ellipsoid and the neutron absorption when the crystal is turned in the scattering plane. Also, the sensitivity is limited, as a factor of 2 in anisotropy only gives a $\sim$30\% intensity change. Taking these effects into account, the data of Ref. \cite{servant02} are consistent with an anisotropy of 2, as found with the more direct and precise method of polarized neutrons used in this work. 

An anisotropy $\chi_c/\chi_a$ of 2 is in reasonable agreement with the factor of 3 inferred from NMR measurements \cite{ishida},  which relies on assumptions on the hyperfine couplings. 
The origin of the anisotropy can be understood in terms of a simple tight-binding model including spin-orbit coupling  \cite{ng}.  

\begin{figure} 
\centering
\includegraphics[width=70mm]{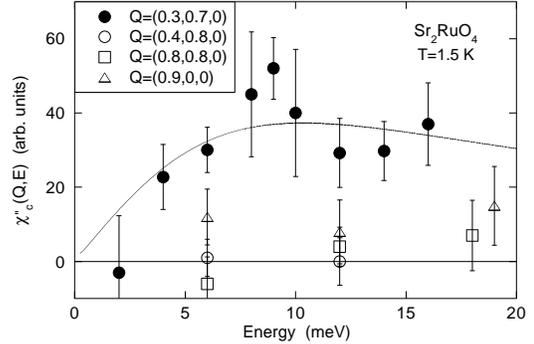}
\caption{\small Energy dependence of the magnetic scattering $\chi"_c(${\bf Q}$,E)$ at different wave vectors {\bf Q} measured with polarized neutrons.}
\label{Edep}
\end{figure}
The energy dependence of the spin fluctuations from unpolarized neutron measurements \cite{sidis,servant00,servant02,braden} shows a quasielastic signal with a characteristic energy of $\Gamma\!\sim$9 meV. The high energy scale makes it difficult to separate the magnetic scattering from phonon scattering processes. The present study using polarized neutrons avoids this problem by measuring directly the magnetic part of the scattering. The resulting $\chi_c$ at the IC position {\bf Q}=(0.3,0.7,0) is shown in Fig.~\ref{Edep}. It is consistent with a quasielastic contribution with an energy scale of $\Gamma$=10$\pm$2 meV. Figure \ref{Edep} also shows data taken at wave vectors away from the IC positions {\bf Q}={\bf k}$_0$. Clearly, there is no substantial magnetic scattering  away from {\bf k}$_0$ in the energy range 4--20 meV, neither in the tail of the IC peaks (cf. also Fig.~\ref{Qdep}) nor at positions where nearly ferromagnetic fluctuations are probed, {\bf Q}=(0.9,0,0). 

The  {\bf q}-independent contribution to the dynamical susceptibility revealed by NMR measurements have not yet been observed  by neutron scattering \cite{sidis,servant00,servant02,braden}. These spin fluctuations, which persists up to temperatures $\sim$500 K, have a characteristic energy of about 50 meV \cite{ishida}. Such fluctuations would be seen in the present study (limitated to $\leq$20 meV due to kinematical constraints) as an initial slope in the energy scan at {\bf Q}=(0.4,0.8,0) shown in Fig. \ref{Edep}. This signal could be close to the limit of the present experimental sensitivity achieved using polarized neutrons. 

We will now discuss the relevance of the present results for the unconventional superconductivity of \Sr2RuO4. The IC fluctuations are clearly anisotropic, as now shown by two different microscopic probes: polarized neutron scattering and NMR.  
In the simplest picture of magnetically mediated superconductivity, IC fluctuations, which are antiferromagnetic in nature rather than ferromagnetic, are believed to promote singlet pairing \cite{monthoux}. However, it has been shown theoretically that {\it anisotropic} fluctuations can favor triplet pairing \cite{sato,kuwabara}.  In such models, an anisotropy $\chi_c/\chi_a\!>$1 will favor  the  most commonly admitted order parameter for describing the superconductivity of \Sr2RuO4. This order parameter has the orbital moment of the pairs lying along the $c$ axis and the spins in the basal plane  (see, e.g., Fig. 20 in Ref. \cite{mackeno}). 

The IC fluctuations occur in the nested  $\alpha$ and $\beta$ bands, 
the so-called  passive bands for superconductivity. However, the superconducting properties of \Sr2RuO4\ are characterstic of the $\gamma$ band, the so-called active band \cite{riseman}. 
The relevance of the IC fluctuations to the superconductivity then obviously depends of the coupling between the different bands \cite{zhito}.

We acknowledge L.P. Regnault for advice on the optimal use of IN22.


\begin{thebibliography}{99}

\bibitem{mackeno} A.P. Mackenzie and Y. Maeno,
Rev. Mod. Phys. 75 (2003) 657. 

\bibitem{sidis} Y. Sidis,
M. Braden, P. Bourges, B. Hennion, S. NishiZaki, Y. Maeno and Y. Mori,
Phys. Rev. Lett. 83 (1999) 320.

\bibitem{servant00} F. Servant,
S. Raymond, B. F\aa k, P. Lejay and J. Flouquet, 
Solid State Commun. 116 (2000) 489.  

\bibitem{servant02} F. Servant,
B. F\aa k, S. Raymond, J.P. Brison, P. Lejay and J. Flouquet, 
Phys. Rev. B 65 (2002) 184511. 
 
\bibitem{braden} M. Braden,
Y. Sidis, P. Bourges, P. Pfeuty, J. Kulda, Z. Mao and Y. Maeno,
Phys. Rev. B 66 (2002) 064522. 

\bibitem{ishida} K. Ishida,
H. Mukuda, Y. Minami, Y. Kitaoka, Z.Q. Mao, H. Fukazawa and Y. Maeno,
Phys. Rev. B 64 (2001) 100501. 

\bibitem{ng} K. Ng and M. Sigrist, 
J. Phys. Soc. Japan 69 (2000) 3764.

\bibitem{monthoux} P. Monthoux and G.G. Lonzarich, 
Phys. Rev. B 59 (1999) 14598.

\bibitem{sato} M. Sato and M. Kohmoto, 
J. Phys. Soc. Japan 69 (2000) 3505.

\bibitem{kuwabara} T. Kuwabara and M. Ogata, 
Phys. Rev. Lett. 85 (2000) 4586.

\bibitem{riseman} T.M. Riseman,
P.G. Kealey, E.M. Forgan, A.P. Mackenzie, L.M. Galvin, A.W. Tyler, S.L. Lee, C.Ager, D. McK. Paul, C.M. Aergerter, R. Cubitt, Z.Q. Mao, T. Akima, and Y. Maeno,
Nature 396 (1998) 242. 

\bibitem{zhito} M.E. Zhitomirsky and T.M. Rice, 
Phys. Rev. Lett. 87 (2001) 0507001.

\end{thebibliography}
\end{document}